\documentclass[seceq]{ptptex}
\title{ Instanton Equations for the Supersymmetric $CP^{N-1}$ Sigma Model\\ on Non(anti)commutative Superspace
}

\author{%       %Use \scshape  for the family name
Kazutoshi Araki$^{1,}$\footnote{E-mail: araki@phys.chuo-u.ac.jp}, 
Takeo Inami$^{1,}$\footnote{E-mail: inami@phys.chuo-u.ac.jp} and 
Hiroaki Nakajima$^{2,}$\footnote{E-mail: nakajima@skku.edu} 
}

\inst{%     %Affiliation, neglected when [addenda] or [errata]
$^1$Department of Physics, Chuo University,\\ Kasuga, Bunkyo-ku, Tokyo, 112-8551, Japan.\\
$^2$Department of Physics and Institute of Basic Science,
Sungkyunkwan University, Suwon 440-746, Korea.
}

\abst{%         %this abstract is neglected when [addenda] or [errata]
We study the instanton equation of the supersymmetric $CP^{N-1}$ sigma model on
non(anti)commutative superspace in two dimensions. We show that the undeformed 
instanton equation is consistent with the deformed equations of motion. Then we
conclude that the instanton equation is not deformed by superspace 
non(anti)commutativity. 
}

\begin{document}

\maketitle

\section{Introduction}
Studies of strings with non-trivial classical values of its field components have  brought about new development in quantum field theories. If one allows some of the field components to posses background values on D-branes, one obtains new classes of quantum field theories in the low energy limits of string theory. They have novel features due to the non(anti)commutativity (NC) properties of spacetime coordinates or superspace coordinates. Superstring theory with its graviphoton field in the R-R sector having a self-dual background value gives rise to supersymmetric Yang-Mills theories with $\mathcal{N}=1/2$ supersymmetry as an effective field theory \cite{seiberg}, providing a good realization of the NC superspace considered in different contexts some time ago \cite{d-susy}.

     NC field theories and SUSY field theories on NC superspace (field theories with deformed SUSY) have properties quite different from the usual field theories, and they need to be examined, in comparison with their commutative partners. Perturbative properties of SUSY field theories on NC superspace have been studied extensively in the last few years. Positive results have been obtained concerning the renormalizability of the model 
\cite{NCWZ,NCYM,Araki:2005nn}.

     One non-pertubative issue in deformed SUSY field theories is whether and how its instanton equation is modified from that of the usual SUSY field theories. This question has been studied in the case of super Yang-Mills theories in d=4, and the modified self-dual equation has been derived \cite{Im}, and is extended to the NC deformed $\mathcal{N}=2$ SUSY case \cite{ItNa}. The solutions to this new self-dual equation have also been studied \cite{SO}.

     $\mathcal{N}=2$ SUSY $CP^{N-1}$ models in d=2 share many nice properties of super Yang-Mills theories and it allows exact treatment. In view of this, we have constructed extension of the SUSY $CP^{N-1}$ model to NC superspace \cite{Ina-Naka,Araki:2005nn}. The usual $CP^{N-1}$ model has a few interesting properties: It is integrable classically \cite{Curtright:1979am} and its multi-instanton solutions are known. It is of interest to see whether these nice features are preserved after extension to NC superspace.

     In this letter we wish to examine the question of whether and how the instanton equation is modified when we extend $\mathcal{N}=2$ SUSY $CP^{N-1}$ model in d=2 to NC superspace. In Yang-Mills theory, construction of multi-instaton solutions and integrability of its self-dual sector are apparently related \cite{Ward:1985gz}. It is of interest to look into the integrability question of the deformed SUSY $CP^{N-1}$ model from this view point. This question will be studied in the forth coming paper.

\section{Instanton equation for $\mathcal{N}=1/2$ Super Yang-Mills Theory}

To the purpose of studying how the instanton equation may be modified in the deformed $\mathcal{N}=2$ SUSY $CP^{N-1}$ model in d=2, it is useful to look at the  modified instanton equation in the deformed super Yang-Mills theory. We begin by summarizing the result of the latter case \cite{Im}.

In four dimensions, the non(anti)commutativity is introduced to $\mathcal{N}$=$1$ superspace by deforming the commutation relations for superspace coordinates $(x^{\mu},\theta^{\alpha},\bar{\theta}^{\dot{\alpha}})$ as \cite{seiberg}
\begin{eqnarray}
&&\{\theta^{\alpha} , \theta^{\beta}\}=C^{\alpha\beta},
\quad \{\theta^{\alpha} , \bar{\theta}^{\dot{\alpha}}\}=\{\bar{\theta}^{\dot{\alpha}} , \bar{\theta}^{\dot{\beta}}\}=0 , \\
&&[ y^{\mu}, y^{\nu}] = 0 ,\quad [ y^{\mu},\theta^{\alpha} ] = [ y^{\mu},\bar{\theta}^{\dot{\alpha}} ] = 0 ,\quad  y^{\mu} = x^{\mu} + i \theta {\sigma}^{\mu} \bar{\theta} .\nonumber 
\end{eqnarray}
The Lagrangian of $\mathcal{N}=1/2$ super Yang-Mills theory is then given by
\begin{eqnarray}
 {\cal L}_{\rm \mathcal{N}=\frac{1}{2}YM} &=& i \tau \int d^2 \theta {\rm tr}W^{\alpha}W_{\alpha} - i
  \bar{\tau}\int d^2 \bar{\theta} \bar{W}^{\dot{\alpha}} \bar{W}_{\dot{\alpha}} \nonumber \\
 & & + (i \tau - i\bar{\tau}) \Bigl( -i C^{\mu \nu} {\rm tr}F_{\mu \nu}
  \bar{\lambda}\bar{\lambda} + \frac{|C|^2}{4} {\rm tr} (\bar{\lambda}\bar{\lambda})^2  \Bigr) ,\label{NCYML}
 \end{eqnarray} 
where $C^{\mu \nu} = C^{\alpha \beta} \epsilon_{\beta \gamma} (\sigma^{\mu
 \nu} )_{\alpha}^{\gamma}$, $|C|^2 = C^{\mu
 \nu} C_{\mu \nu} . $
$W_{\alpha}$ and $\bar{W}_{\dot{\alpha}}$ are the field strength superfield. $\tau$ and $\bar{\tau}$ are the gauge coupling constants.
The above Lagrangian in the ordinary sense is invariant under the following $\mathcal{N}=1/2$ supersymmetry transformations.
\begin{eqnarray}
&&\delta \lambda = i \epsilon D + \sigma^{\mu \nu}\epsilon\Bigl( F_{\mu
 \nu} + \frac{i}{2} C_{\mu \nu}  \bar{\lambda}\bar{\lambda}  \Bigr) \label{eq:vsusy} ,   \\
&& \delta A_{\mu} = -i \bar{\lambda} \bar{\sigma}^{\mu}\epsilon , \hspace{5mm} 
\delta D = - \epsilon \sigma^{\mu} D_{\mu} \bar{\lambda} ,  \hspace{5mm}
\delta \bar{\lambda} = 0 \hspace{0.1cm}. \nonumber
\end{eqnarray}
The other half $\bar{Q}$ of $\mathcal{N}=1$ SUSY is broken.

The usual $\mathcal{N}=1$ super Yang-Mills theory possesses instantons, which are solutions to the set of anti-self-dual equations,
\begin{eqnarray}
F_{\mu \nu}^{+} = 0 , \hspace{7mm}
{\sigma}^{\mu} D_\mu \bar{\lambda} =  0 , \hspace{7mm}
\lambda =  0 .\label{eq:OGSD}
\end{eqnarray}
They also satisfy the equation of motion. After extension of the super Yang-Mills theory to the NC superspace, we may define instantons as solutions to eq. (\ref{eq:OGSD}). However, they are not solutions of the equation of motion any more. This apparent contradiction should be resolved by modifying the instanton equation for the super Yang-Mills theory.

The correct modification was found by Imaanpur \cite{Im} and Britto et al \cite{SO} by considering the BPS condition. It reads as follows. \\
%Instanton solutions satisfy equation of motion, and saturate BPS bound. From eq%s. (\ref{eq:vsusy}), deformed instanton equation can be considered the followin%g,  
The anti-self-dual equation is deformed and reads as
\begin{eqnarray}
F_{\mu \nu}^{+} + \frac{i}{2} C_{\mu \nu}  \bar{\lambda}\bar{\lambda} =
  0 , \hspace{7mm}
{\sigma}^{\mu} D_\mu \bar{\lambda} =  0 , \hspace{7mm}
\lambda =  0 .\label{eq:GSD}
\end{eqnarray}
The self-dual equation is not modified:
\begin{eqnarray}
F_{\mu \nu}^{-}  =  0 , \hspace{7mm}
\bar{\sigma}^{\mu} D_\mu \lambda =  0 , \hspace{7mm}
\bar{\lambda} =  0 .\label{eq:GASD}
\end{eqnarray}

It is easy to see that solutions to eqs. (\ref{eq:GSD}) or (\ref{eq:GASD}) are also solutions to the equations of motion, which are given by
\begin{eqnarray}
D^\mu ( F_{\mu \nu} + i C_{\mu \nu}  \bar{\lambda}\bar{\lambda}) &=&
 D^\mu \Bigl(  F_{\mu \nu}^{-} + \frac{i}{2} C_{\mu \nu}
 \bar{\lambda}\bar{\lambda}  \Bigr) \nonumber \\
&=& D^\mu \Bigl(  F_{\mu \nu}^{+} + \frac{i}{2} C_{\mu \nu}
 \bar{\lambda}\bar{\lambda}  \Bigr) = 0 \hspace{0.1cm}, \\
\bar{\sigma}^{\mu} D_\mu \lambda &=& -C^{\mu\nu} \bar{\lambda} \Bigl(  F_{\mu
 \nu}^{+} + \frac{i}{2} C_{\mu \nu} (\bar{\lambda}\bar{\lambda})  \Bigr) , \nonumber \\
{\sigma}^{\mu} D_\mu \bar{\lambda} &=&  0 \hspace{0.1cm} . \nonumber
\end{eqnarray}

\section{Lagrangian of the $CP^{N-1}$ model on NC superspace}
SUSY $CP^{N-1}$ model in $d=2$ can be obtained
from that in $d=4$ by dimensional reduction \cite{Aoyama:1980yw}. The same method can be used to obtain the $CP^{N-1}$ model on NC superspace in $d=2$ \cite{Ina-Naka}.

We denote the sets of scalar and Dirac fields by $\varphi^{a}$, $\bar{\varphi}^{\bar{a}}$ and $\chi^{a}$, $\bar{\chi}^{\bar{a}}$ $(a=1,2,\cdots,N-1)$ respectively. The Lagrangian of the $CP^{N-1}$ model on NC superspace in $d=2$ is written in terms of the component fields as
\begin{eqnarray}
\mathcal{L} &=& \mathcal{L}_0 + \mathcal{L}_C , \\
\mathcal{L}_0 &=& g_{a\bar{b}}\partial_{\mu}\varphi^{a}\partial^{\mu}\bar{\varphi}^{\bar{b}} + i g_{a\bar{b}}\bar{\chi}^{\bar{b}} \bar{\sigma}^{\mu} D_{\mu}\chi^{a} - \frac{1}{8} R_{a\bar{b}c\bar{d}}(\bar{\chi}^{\bar{b}} \bar{\sigma}^{\mu} \chi^{a} )(\bar{\chi}^{\bar{d}} \bar{\sigma}_{\mu} \chi^{c} ) , \label{L0}\\
\mathcal{L}_C &=& g_{a\bar{b}} g_{c\bar{d}} ( C^{11} \chi^{a}_{+} \chi^{c}_{+} - C^{22} \chi^{a}_{-} \chi^{c}_{-} ) \epsilon^{\mu \nu} (\partial_{\mu}\bar{\varphi}^{\bar{b}})(\partial_{\nu}\bar{\varphi}^{\bar{d}}) .\label{LC}
\end{eqnarray}
Here $\chi^{a}_{+},\chi^{a}_{-}$ are the two components of the 2D spinor $\chi^{a}$. $\mathcal{L}_0$ is the undeformed part, namely, the usual SUSY 
$CP^{N-1}$ Lagrangian \cite{D'Adda:1978kp,Cremmer:1978bh}. $\mathcal{L}_C$ is the new term due to superspace
non(anti)commutativity.
$g_{a\bar{b}}$ is the Fubini-Study metric on $CP^{N-1}$. $\Gamma^{a}_{bc}$ and $R_{a\bar{b}c\bar{d}}$ are the Christoffel symbol and the Riemann curvature tensor, respectively. $D_{\mu}\chi^{a}$ is the covariant derivative. They are given by
\begin{eqnarray}
g_{a\bar{b}} &=& \frac{(1 + \bar{\varphi} \varphi ) \delta_{a\bar{b}} - \bar{\varphi}_{a} \varphi_{\bar{b}}}{(1 + \bar{\varphi}\varphi)^2} ,\\
\Gamma^{a}_{bc} &=& g^{a\bar{d}}\partial_{b}g_{c\bar{d}},\\ 
D_{\mu}\chi^{a} &=& \partial_{\mu}\chi^{a} + 
 \Gamma^{a}_{bc}(\partial_{\mu}\varphi^{b})\chi^{c},\\
R_{a\bar{b}c\bar{d}} &=& -g_{a\bar{e}}\partial_{c}(g^{f\bar{e}}\partial_{\bar{d}}g_{f\bar{b}})=g_{a\bar{b}}g_{c\bar{d}}+g_{a\bar{d}}g_{c\bar{b}}.
\end{eqnarray}

\section{Instanton equation for $CP^{N-1}$ model on NC superspace}

We introduce the complex coordinates
\begin{eqnarray}
z = x^0 + i x^1 ,\quad \bar{z} = x^0 - i x^1 .  \label{eq:comp}
\end{eqnarray}
We write $\partial = \partial_{z}$ and $\bar{\partial} = \partial_{\bar{z}}$. The antisymmetric tensor $\epsilon^{\mu \nu}$ is written as
\begin{eqnarray}
%\epsilon_{\mu \nu} &\to& \epsilon_{z z} = \epsilon_{\bar{z} \bar{z}} = 0 ,\quad \epsilon_{z \bar{z}} = \frac{i}{2} ,\quad \epsilon_{\bar{z} z} = - \frac{i}{2} \\
\epsilon^{z z} = \epsilon^{\bar{z} \bar{z}} = 0 ,\quad \epsilon^{z \bar{z}} = -2i ,\quad \epsilon^{\bar{z} z} = 2i .  \label{eq:com-eps}
\end{eqnarray}
The original self-dual equation and anti-self-dual equation can be written in terms of the complex coordinates (\ref{eq:comp}) as follows.  \\
Self-dual equation:
\begin{eqnarray}
\bar{\partial} \varphi^a = \partial \bar{\varphi}^{\bar{a}} = 0 ,  \hspace{5mm}
D_{z} \chi^{a}_{-} = D_{\bar{z}} \bar{\chi}^{\bar{a}}_{+} = 0 ,  \hspace{5mm}
\chi^{a}_{+} = \bar{\chi}^{\bar{a}}_{-} = 0  .\label{SD}
\end{eqnarray}
Anti-self-dual equation:
\begin{eqnarray}
\partial \varphi^{a} = \bar{\partial} \bar{\varphi}^{\bar{a}} = 0 , \hspace{5mm}
D_{\bar{z}} \chi^{a}_{+} = D_{z} \bar{\chi}^{\bar{a}}_{-} = 0 , \hspace{5mm}
\chi^{a}_{-} = \bar{\chi}^{\bar{a}}_{+} = 0  .\label{ASD}
\end{eqnarray}

We now study the equations of motion. Let $\Omega $ be one of the fields $\varphi^{a}$, $\bar{\varphi}^{\bar{a}}$, $\chi^{a}_{+}$, $\chi^{a}_{-}$,$\bar{\chi}^{\bar{a}}_{+}$, $\bar{\chi}^{\bar{a}}_{-} $. The equation of motion for $CP^{N-1}$ model on NC superspace is written as follows 
\begin{equation}
\partial_{\mu} \frac{\partial \mathcal{L}}{\partial ( \partial_{\mu} \Omega  )} - \frac{\partial \mathcal{L}}{\partial \Omega  } = \Bigl( \partial_{\mu} \frac{\partial \mathcal{L}_0}{\partial ( \partial_{\mu} \Omega  )} - \frac{\partial \mathcal{L}_0}{\partial \Omega  } \Bigr) + \Bigl( \partial_{\mu} \frac{\partial \mathcal{L}_C}{\partial ( \partial_{\mu} \Omega  )} - \frac{\partial \mathcal{L}_C}{\partial \Omega  } \Bigr) = 0 . \label{L0-EOM_all}
\end{equation}
The instanton solution before the deformation is a special solution for the equation of motion of $\mathcal{L}_0$ part. Then we have
\begin{eqnarray}
\Bigl( \partial_{\mu} \frac{\partial \mathcal{L}_0}{\partial ( \partial_{\mu} \Omega  )} - \frac{\partial \mathcal{L}_0}{\partial \Omega  } \Bigr) \Bigr|_{\mbox{instanton}} &=& 0 . \label{L0-EOM1}
\end{eqnarray}

\section{Equation of motion of $\mathcal{L}_C$ Part}

We now deal with the second term in eq. (\ref{L0-EOM_all}). The factor of derivative of $\bar{\varphi}$ in $\mathcal{L}_C$ is

\begin{eqnarray}
\epsilon^{\mu \nu} (\partial_{\mu}\bar{\varphi}^{\bar{b}})(\partial_{\nu}\bar{\varphi}^{\bar{d}}) = 2i ( \bar{\partial} \bar{\varphi}^{\bar{b}} \partial \bar{\varphi}^{\bar{d}} - \partial \bar{\varphi}^{\bar{b}} \bar{\partial} \bar{\varphi}^{\bar{d}} ) .
\end{eqnarray}
$\mathcal{L}_C$ can be rewritten as
\begin{eqnarray}
\mathcal{L}_C &=& \sum_{F=+,-} C^{F} g_{a\bar{b}} g_{c\bar{d}} \chi^{a}_{F} \chi^{c}_{F}  \bar{\partial} \bar{\varphi}^{\bar{b}} \partial \bar{\varphi}^{\bar{d}}  ,
\end{eqnarray}
where we have set $C^{+} = 4i C^{11}$, $C^{-} = - 4i C^{22}$. We analyse the equations of motion for each of the scalar and fermion fields, $\varphi^{a}$, $\bar{\varphi}^{\bar{a}}$, $\chi^{a}_{+}$, $\chi^{a}_{-}$,$\bar{\chi}^{\bar{a}}_{+}$, $\bar{\chi}^{\bar{a}}_{-} $, which we denote by $\Omega$ collectively. \\
We first consider the case of $\Omega =\bar{\varphi}^{\bar{a}}$.
\begin{eqnarray}
\partial \frac{\partial \mathcal{L}_C}{\partial (\partial \bar{\varphi}^{\bar{e}})} & = & \sum_{F=+,-} C^{F} \chi^{a}_{F} \chi^{c}_{F} \Bigl[ g_{a\bar{b}} g_{c\bar{e}} (\partial \bar{\partial} \bar{\varphi}^{\bar{b}} )  + ( g_{c\bar{e}} \bar{\partial}_{\bar{l}} g_{a\bar{b}} + g_{a\bar{b}} \bar{\partial}_{\bar{l}} g_{c\bar{e}} )  ( \partial \bar{\varphi}^{\bar{l}} ) ( \bar{\partial} \bar{\varphi}^{\bar{b}} ) \Bigr]  \nonumber \\
& & + \sum_{F=+,-} C^{F} g_{a\bar{b}} g_{c\bar{e}} \Bigl[ ( \partial \chi^{a}_{F} + \partial \varphi^l \cdot \Gamma^a_{lm} \chi^{m}_{F} ) \chi^{c}_{F}   \nonumber \\
& & \quad \quad \quad \quad \quad \quad \quad \quad \quad + \chi^{a}_{F} ( \partial  \chi^{c}_{F} + \partial \varphi^l \cdot \Gamma^c_{lm} \chi^{m}_{F} ) \Bigr] ( \bar{\partial} \bar{\varphi}^{\bar{b}} ),  \label{LC-EOM1}
\end{eqnarray}
\begin{eqnarray}
\bar{\partial} \frac{\partial \mathcal{L}_C}{\partial ( \bar{\partial} \bar{\varphi}^{\bar{e}})} 
& = & \sum_{F=+,-} C^{F} \chi^{a}_{F} \chi^{c}_{F} \Bigl[ g_{a\bar{e}} g_{c\bar{d}} ( \bar{\partial} \partial \bar{\varphi}^{\bar{d}} ) + ( g_{c\bar{d}} \bar{\partial}_{\bar{l}} g_{a\bar{e}} + g_{a\bar{e}} \bar{\partial}_{\bar{l}} g_{c\bar{d}} )  ( \bar{\partial} \bar{\varphi}^{\bar{l}} ) ( \partial \bar{\varphi}^{\bar{d}} ) \Bigr]  \nonumber \\
& & + \sum_{F=+,-} C^{F} g_{a\bar{e}} g_{c\bar{d}} \Bigl[ ( \bar{\partial} \chi^{a}_{F} + \bar{\partial} \varphi^l \cdot \Gamma^a_{lm} \chi^{m}_{F} ) \chi^{c}_{F}   \nonumber \\
& & \quad \quad \quad \quad \quad \quad \quad \quad \quad + \chi^{a}_{F} ( \bar{\partial}  \chi^{c}_{F} + \bar{\partial} \varphi^l \cdot \Gamma^c_{lm} \chi^{m}_{F} ) \Bigr] ( \partial \bar{\varphi}^{\bar{d}} ),
\end{eqnarray}
\begin{eqnarray}
\frac{\partial \mathcal{L}_C}{\partial \bar{\varphi}^{\bar{e}} } & = & \sum_{F=+,-} C^{F} \bar{\partial}_{\bar{e}} ( g_{a\bar{b}} g_{c\bar{d}} ) \cdot \chi^{a}_{F} \chi^{c}_{F}  ( \bar{\partial} \bar{\varphi}^{\bar{b}} )( \partial \bar{\varphi}^{\bar{d}} ) . \label{LC-EOM-OO}
\end{eqnarray}
From eqs. (\ref{LC-EOM1}) - (\ref{LC-EOM-OO}) and after changes of dummy indices, we obtain
\begin{eqnarray}
\partial_{\mu} \frac{\partial \mathcal{L}_C}{\partial ( \partial_{\mu} \bar{\varphi}^{\bar{e}}  )} - \frac{\partial \mathcal{L}_C}{\partial \bar{\varphi}^{\bar{e}}  } & = & \sum_{F=+,-} C^{F} \chi^{a}_{F} \chi^{c}_{F} \Bigl[ \bar{\partial}_{\bar{l}} ( g_{a\bar{b}} g_{c\bar{e}} ) + \bar{\partial}_{\bar{b}} ( g_{a\bar{e}}  g_{c\bar{l}})  \label{LC-EOM-OOO}  \\
& & \hspace{48mm} - \bar{\partial}_{\bar{e}} ( g_{a\bar{b}} g_{c\bar{l}} ) \Bigr] \underline{( \partial \bar{\varphi}^{\bar{l}} )} ( \bar{\partial} \bar{\varphi}^{\bar{b}} )   \nonumber \\
& & + ( g_{a\bar{b}} g_{c\bar{e}} - g_{a\bar{e}} g_{c\bar{b}} ) \Bigl[ C_+ D_{z} \chi^{a}_{+} \cdot \underline{\chi^{c}_{+}} \bar{\partial} \bar{\varphi}^{\bar{b}} 
 + C_- \underline{D_{z} \chi^{a}_{-}} \cdot \chi^{c}_{-} \bar{\partial} \bar{\varphi}^{\bar{b}}    \nonumber \\
& & \hspace{30mm} - C_+ D_{\bar{z}} \chi^{a}_{+} \cdot \underline{\chi^{c}_{+}} \underline{\partial \bar{\varphi}^{\bar{b}}} 
 - C_- D_{\bar{z}} \chi^{a}_{-} \cdot \chi^{c}_{-} \underline{\partial \bar{\varphi}^{\bar{b}}}  \Bigr].  \nonumber
\end{eqnarray}
We note that all terms in eq. (\ref{LC-EOM-OOO}) contain factors (they are underlined) which vanish by use of the SD equation (\ref{SD}). Hence, we have derived
\begin{eqnarray}
\Bigl( \partial_{\mu} \frac{\partial \mathcal{L}_C}{\partial ( \partial_{\mu} \bar{\varphi}^{\bar{e}}  )} - \frac{\partial \mathcal{L}_C}{\partial \bar{\varphi}^{\bar{e}}  } \Bigr) \Bigr|_{\mbox{SD eq.}} &=& 0 . \label{L00C-EOM1}
\end{eqnarray}
For the other fields $\Omega =\varphi^{a},\chi^{a}_{+},\chi^{a}_{-},\bar{\chi}^{\bar{a}}_{+},\bar{\chi}^{\bar{a}}_{-} $, $\mathcal{L}_C$ does not contain $\partial_{\mu} \Omega$. Hence, their equations of motion are
\begin{equation}
\partial_{\mu} \frac{\partial \mathcal{L}_C}{\partial ( \partial_{\mu} \Omega  )} - \frac{\partial \mathcal{L}_C}{\partial \Omega  } =  - (\bar{\partial} \bar{\varphi}^{\bar{b}}) \underline{ ( \partial \bar{\varphi}^{\bar{d}} ) } \frac{\partial }{\partial \Omega  } \sum_{F=+,-} C^{F} g_{a\bar{b}} g_{c\bar{d}} \chi^{a}_{F} \chi^{c}_{F}. \label{LC-EOM-O}
\end{equation}
Because of the factor $\partial \bar{\varphi}^{\bar{d}}$, the same equation as (\ref{L00C-EOM1}) holds. \\
We now have shown
\begin{eqnarray}
\Bigl( \partial_{\mu} \frac{\partial \mathcal{L}_C}{\partial ( \partial_{\mu} \Omega  )} - \frac{\partial \mathcal{L}_C}{\partial \Omega  } \Bigr) \Bigr|_{\mbox{SD eq.}} &=& 0  \label{L0C-EOM1}
\end{eqnarray}
for all fields. It means that the original instanton solutions are solutions to the equation of motion of the $CP^{N-1}$ model on NC superspace. The same statement can be made for anti-instantons.

Our result that the {\it undeformed} instanton equation is in accord with the {\it deformed} equation of motion is in clear contrast with the super Yang-Mills case and is surprising at first sight. We recall that the $\mathcal{N}=1/2$ supersymmetry transformation for the NC deformed supersymmetric $CP^{N-1}$ model is not affected by the non(anti)commutativity $C^{\alpha \beta}\neq 0$ \cite{Ina-Naka}, unlike the the super Yang-Mills case, eq. (\ref{eq:OGSD}). The absence of corrections to the instanton equation due to the deformation in the supersymmetric $CP^{N-1}$ model is related to the absence of corrections to the SUSY transformation due to $C^{\alpha \beta}$. This is because the instanton equation is a BPS equation.

%%%%%%%%%%%%%%%%%%%%%%%%%%%%%%%%%%%%%%%%%%%%%%%%%%%%%%%%%%%%%%%%%%%%%%%%%%%%%%%

\section*{Acknowledgements} 
This work is supported partially by  the research grant of Japanese
Ministry of Education and and Science (Kiban B and Kiban C) and Chuo University research grant.
K.A. is supported by the Research Assistant Fellowship of Chuo University. 
This work is also supported by Astrophysical Research
Center for the Structure and Evolution of the Cosmos (ARCSEC).

\end{document}